\documentclass[a4paper,fleqn,usenatbib,useAMS]{mn2e}

\usepackage{graphicx}	
\usepackage{amsmath}	
\usepackage{amssymb}	
\usepackage{multicol}        
\usepackage{bm}		
\usepackage{pdflscape}	
\usepackage[colorlinks]{hyperref}
\usepackage{comment}
\usepackage{subcaption}
\usepackage{overpic,color}
\usepackage{caption}
\usepackage{multicol}        
\usepackage{mathptmx}
\usepackage[T1]{fontenc}
\usepackage{ae,aecompl}
\usepackage{pdflscape}	
\usepackage{booktabs} 
\usepackage{epsfig} 
\usepackage{natbib}
\usepackage{textcomp}
\usepackage[export]{adjustbox}
\usepackage[T1]{fontenc}
\usepackage{aecompl}
\usepackage[justification=centering]{caption}

\newcommand{{\lenstool}}{{\small LENSTOOL}}
\newcommand{{\bayesys}}{{\small BAYESYS}}
\newcommand{{\astrodendro}}{{\small ASTRODENDRO}}
\newcommand{{\clumpfind}}{{\small CLUMPFIND}}

\title[Monthly Notices of the Royal Astronomical Society]
  {Resolving star-forming clumps in a z $\sim$ 2 lensed galaxy: a pixelated Bayesian approach}
\author[Sharma et al.]
  {Soniya Sharma$^{1,2,3}$, Johan Richard$^4$, Tiantian Yuan$^{3,5}$, Vera Patr\'icio$^6$, Lisa Kewley$^{2,3}$,\\
 \newauthor
 Jane R. Rigby$^{1}$, Anshu Gupta$^{7,2}$, Nicha Leethochawalit$^{8}$\\
 $^1$ Observational Cosmology Lab, NASA Goddard Space Flight Center, Greenbelt, MD 20771, USA \\
 $^2$Research School of Astronomy and Astrophysics, Australian National University, Cotter Road, ACT 2611, Australia\\
 $^3$ARC Centre of Excellence for All Sky Astrophysics in 3 Dimensions (ASTRO 3D), Australia\\
 $^4$ Univ Lyon, Univ Lyon1, Ens de Lyon, CNRS, Centre de Recherche Astrophysique de Lyon UMR5574, F-69230, Saint-Genis-Laval, France\\
 $^5$ Centre for Astrophysics and Supercomputing, Swinburne University of Technology, Hawthorn, Victoria 3122, Australia \\
 $^6$ DARK, Niels Bohr Institute, University of Copenhagen, Lyngbyvej 2, 2100 Copenhagen, Denmark \\
 $^7$ School of Physics, University of New South Wales, Kensington, Australia \\
 $^8$ School of Physics, University of Melbourne, Parkville, Victoria, Australia\\
  }
\date{Released 2021Xxxxx XX}

\pagerange{\pageref{firstpage}--\pageref{lastpage}} \pubyear{2021}

\def\LaTeX{L\kern-.36em\raise.3ex\hbox{a}\kern-.15em
    T\kern-.1667em\lower.7ex\hbox{E}\kern-.125emX}

\begin{document}
\label{firstpage}

\maketitle
\begin{abstract}
We present a pixelized source reconstruction method applied on Integral Field Spectroscopic (IFS) observations of gravitationally lensed galaxies. We demonstrate the effectiveness of this method in a case study on the clumpy morphology of a $z \sim 2$ lensed galaxy behind a group-scale lens. We use a Bayesian forward source modelling approach to reconstruct the surface brightness distribution of the source galaxy on a uniformly pixelized grid while accounting for the image point spread function (PSF). The pixelated approach is sensitive to clump sizes down to 100 pc and resolves smaller clump sizes with an improvement in the signal to noise ratio (SNR) by almost a factor of ten compared with more traditional ray-tracing approaches. \\
\noindent
\textit{}
\end{abstract}


\section{Introduction}\label{intro}
An increasingly large number of high-redshift systems lensed by cluster scale and galaxy scale lenses are being studied at the finest angular scales \citep[][]{Sharon12, Dye14, Sharma18,Cheng19}. A plethora of different techniques have enabled the construction of high precision lensing magnification maps of galaxy clusters that act as strong lenses \citep{Johnson14, Meneg17, Diego18, Cibirka18}. However, the accuracy of the source-plane reconstruction does not rely on the lensing mass model alone. The distortion of PSF from the image to the source plane adds a bottleneck in recovering the source-plane features to the highest possible signal-to-noise and spatial resolutions. This bottleneck is most noticeable in seeing-limited observations, such as IFS observations of lensed galaxies. 

Observations of galaxies in the Hubble Space Telescope Ultra Deep Fields, further investigated at higher spatial resolution with strong gravitational lensing have revealed a large number of clumpy irregular disks at higher redshifts (z > 1) \citep{Swinbank06, Elmegreen07,  Jones10, Forster11, Livermore12, Livermore15, Johnson17}. However, the physical sizes of star-forming regions in these galaxies appear to depend on the effective spatial resolution. For the most highly magnified systems, the effects of lensing PSF (projection of observed PSF either from the atmosphere or telescope back to the unlensed plane), present a major hurdle in accurately recovering the intrinsic properties of lensed galaxies.

Reconstruction of giant arcs behind galaxy groups and clusters using a forward approach can overcome the challenges posed by the source plane PSF in modeling extended sources. The aim of a forward approach is simple: for a given lens model, utilize the constraints from the extended surface-brightness profile of the lensed source to reconstruct its intrinsic profile. Various methods have been developed for the inversion of extended lensed images; more so for galaxy-scale lenses \citep{Warren03, Claskens06, Brewer06, Tagore14, Night18, Galan20} compared with the more complex mass distributions of group scale/cluster lenses \citep{Yang20, Suyu06}.

\begin{figure}
\centering
\includegraphics[width=0.95\linewidth]{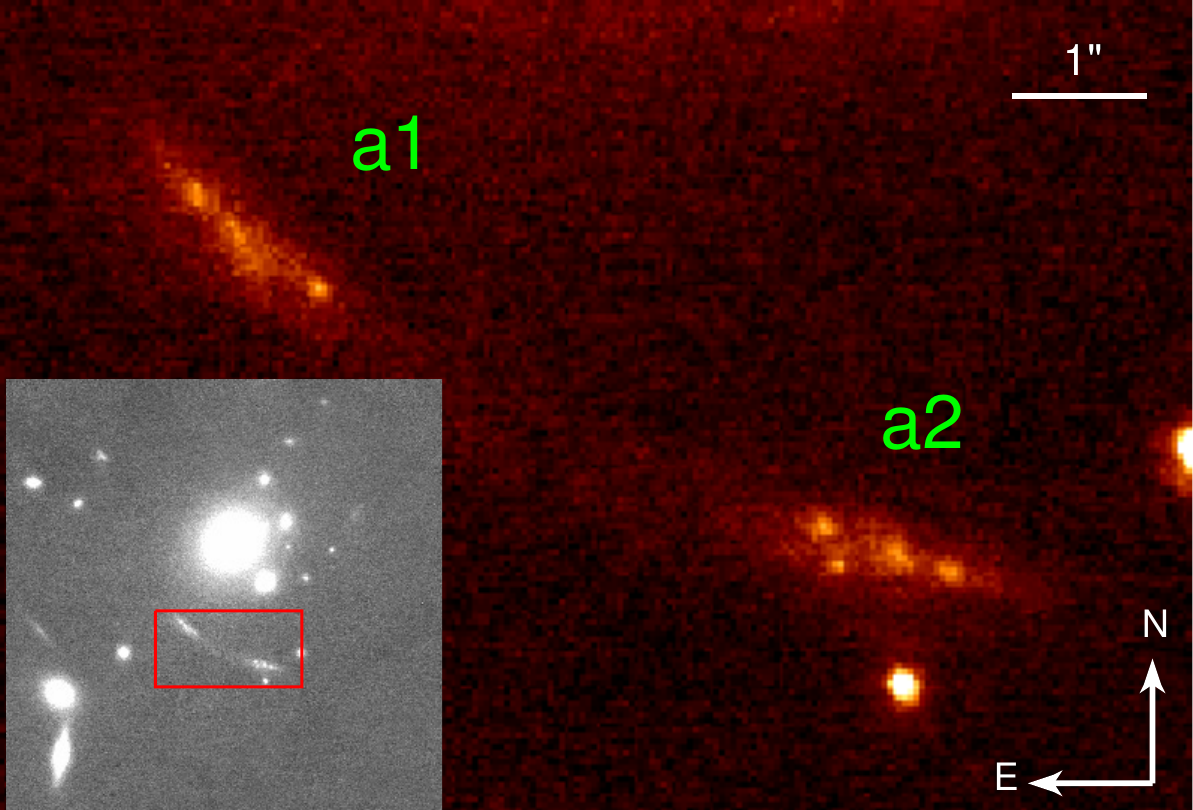}
\caption{High resolution NIRC2 imaging data of cswa128, zoomed in to highlight the two multiple images a1 and a2 at $z= 2.225$. Inset shows the complete Field of View (FOV) of the lensing galaxy group at $z = 0.214$. The spatial resolution is $0. \!\!^{\prime\prime}04$. Please to refer to Paper I for further details on this data.} 
\label{fig:fig1}
\end{figure}

In this letter, we combine a forward approach with a fully automated image inversion technique newly implemented in the publicly available software {\lenstool}\footnote{http://projects.lam.fr/projects/lenstool}\citep{Kneib93, Jullo07} for the source plane reconstruction of strongly lensed galaxies. We describe the first application of this technique on one of the brightest targets from the Cambridge Sloan Survey of Wide Arcs in the Sky \citep[CASSOWARY;][]{Stark13} sample, SDSS1958+5950 (hereafter referred by its survey ID: cswa128; $z = 2.225$). We show that this technique can successfully recover clump sizes with a higher SNR in the source plane as compared to the traditional ray-tracing. 

\begin{figure*}
\centering
\begin{subfigure}[b]{0.50\textwidth}
\centering
\includegraphics[ width=0.99\linewidth,valign=t]{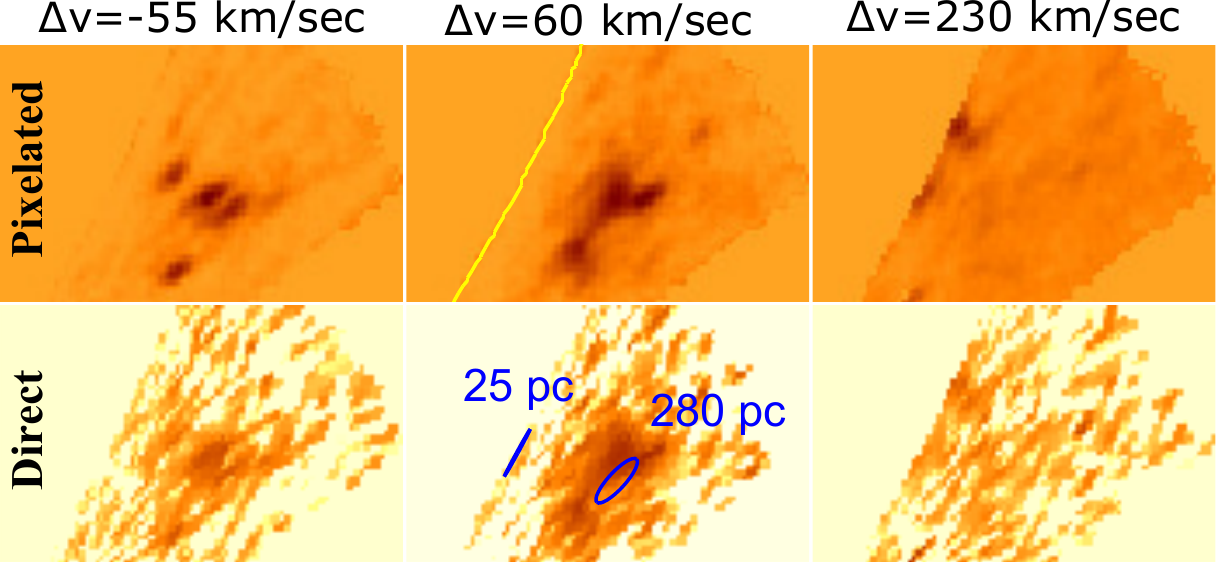}
\caption{}
\label{fig:fig2a}
\end{subfigure}
\begin{subfigure}[b]{0.49\textwidth}
\includegraphics[ trim=1.3cm 0.5cm 0.3cm 2.0cm, width=0.45\linewidth, valign=t]{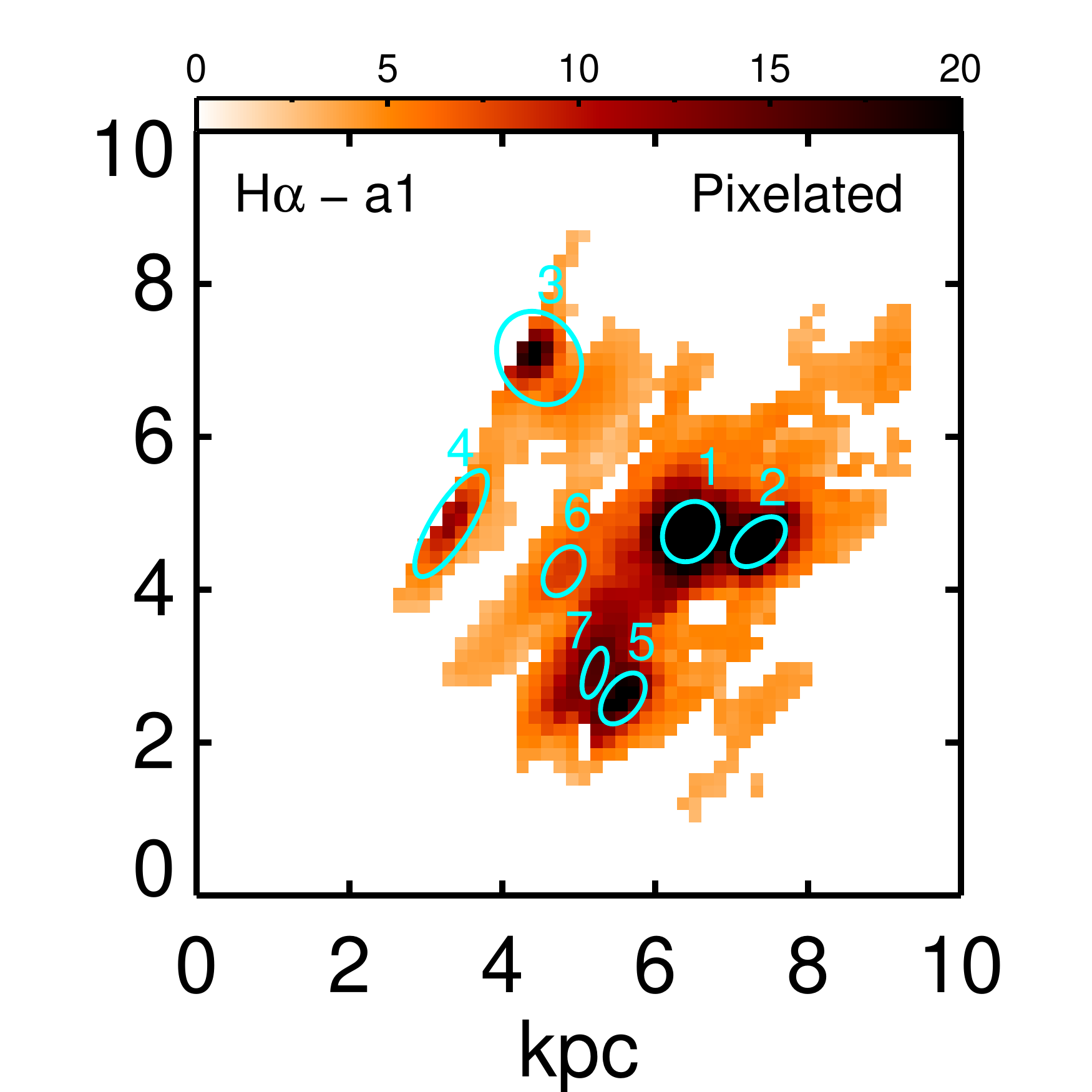}
\includegraphics[ trim=1.3cm 0.5cm 0.3cm 2.0cm, width=0.45\linewidth, valign=t]{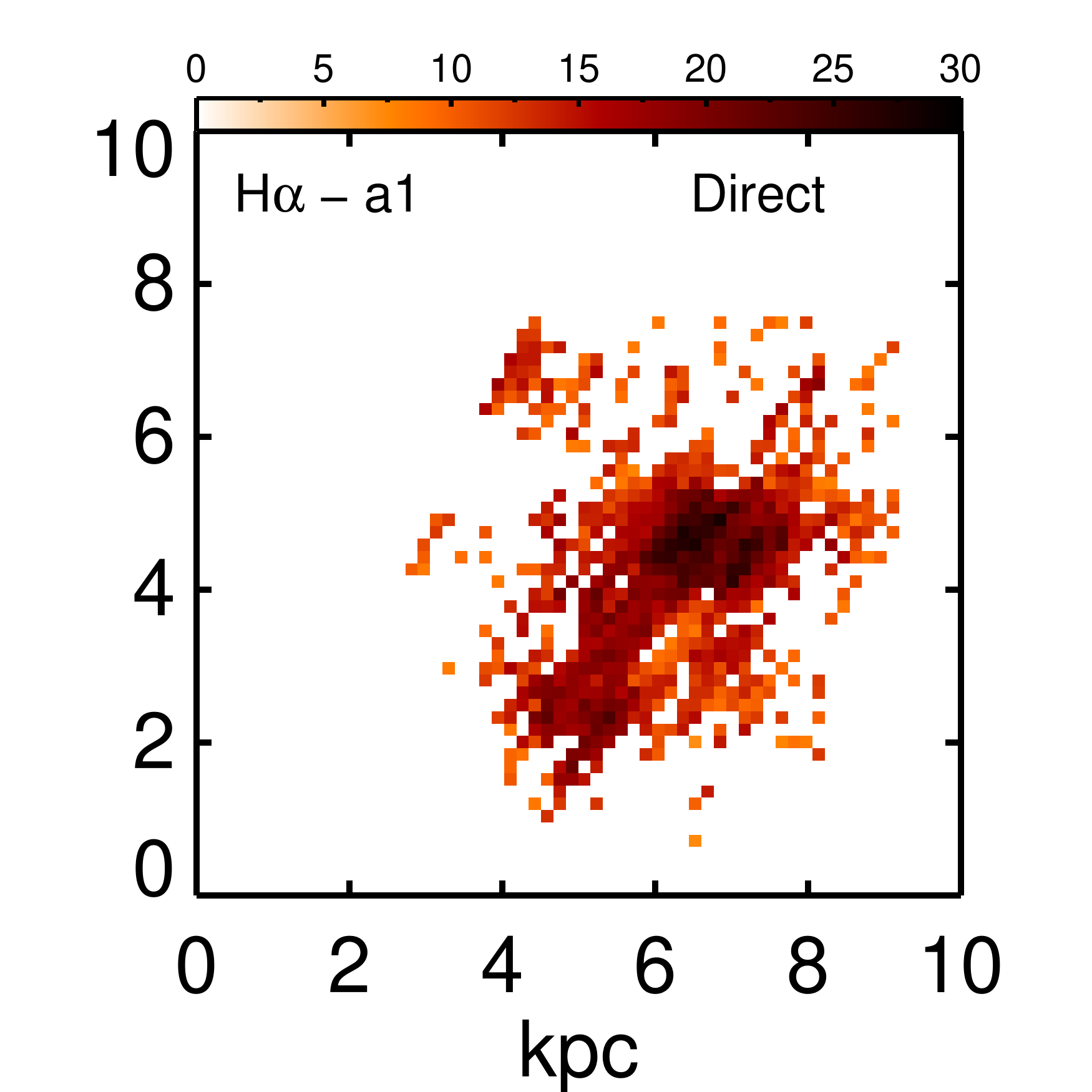}
\caption{}
\label{fig:fig2b}
\end{subfigure}
\begin{subfigure}[b]{0.50\textwidth}
\vspace{0.4cm}
\centering
\includegraphics[ width=0.99\linewidth,valign=t]{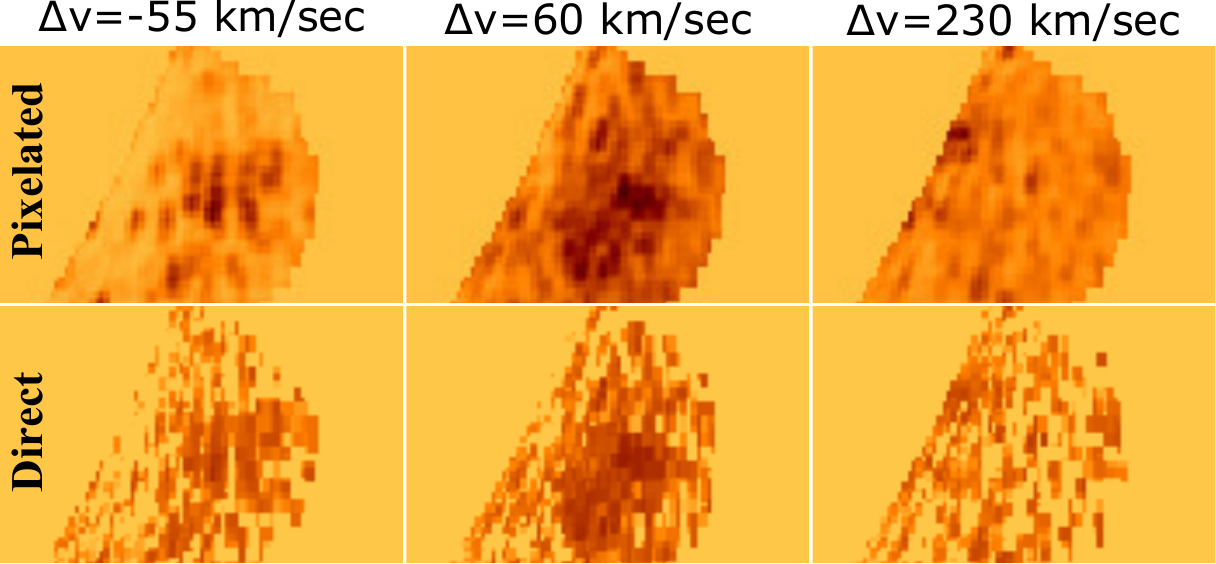}
\caption{}
\label{fig:fig2c}
\end{subfigure}
\begin{subfigure}[b]{0.49\textwidth}
\includegraphics[ trim=1.3cm 0.5cm 0.3cm 2.0cm, width=0.45\linewidth,valign=t]{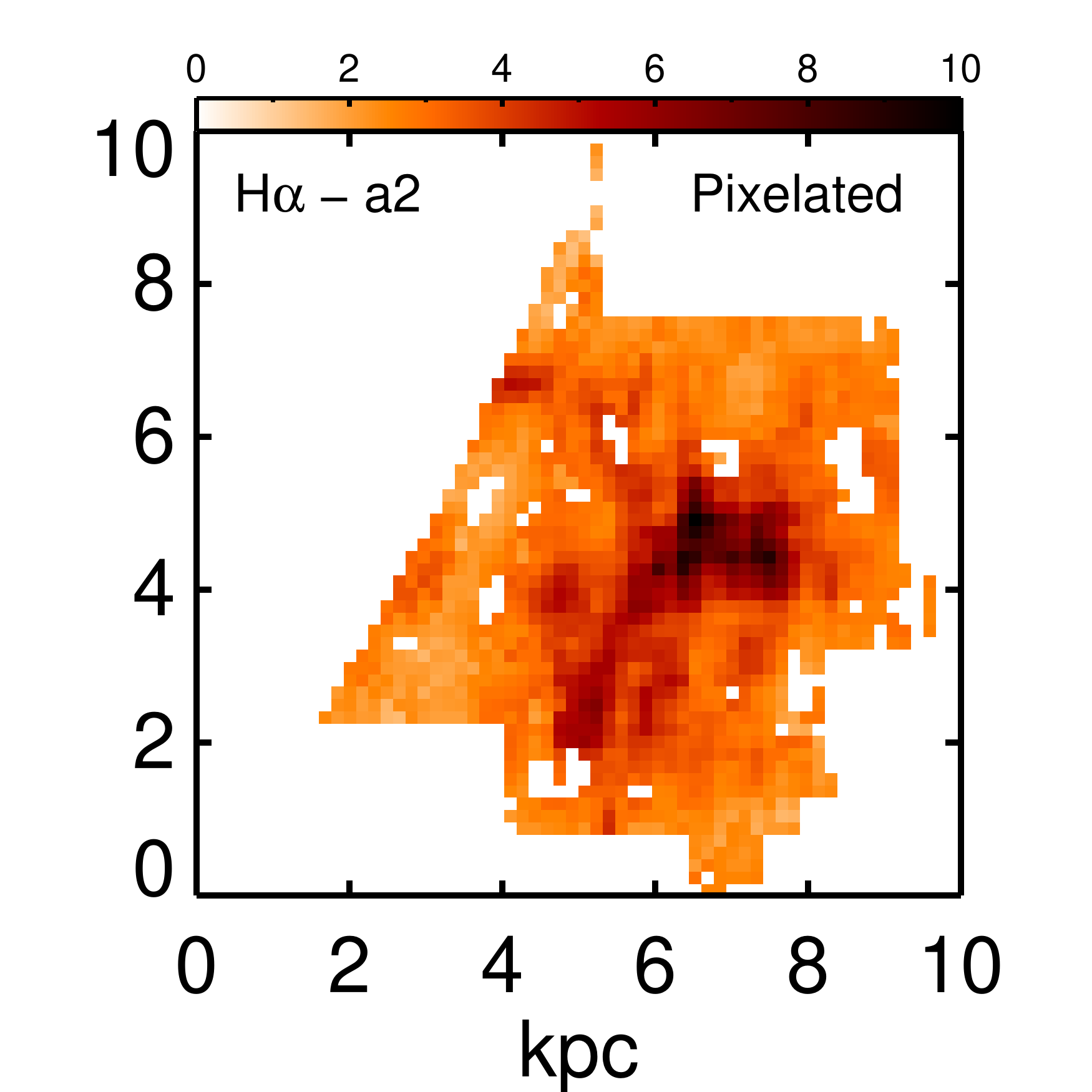}
\includegraphics[ trim=1.3cm 0.5cm 0.3cm 2.0cm, width=0.45\linewidth,valign=t]{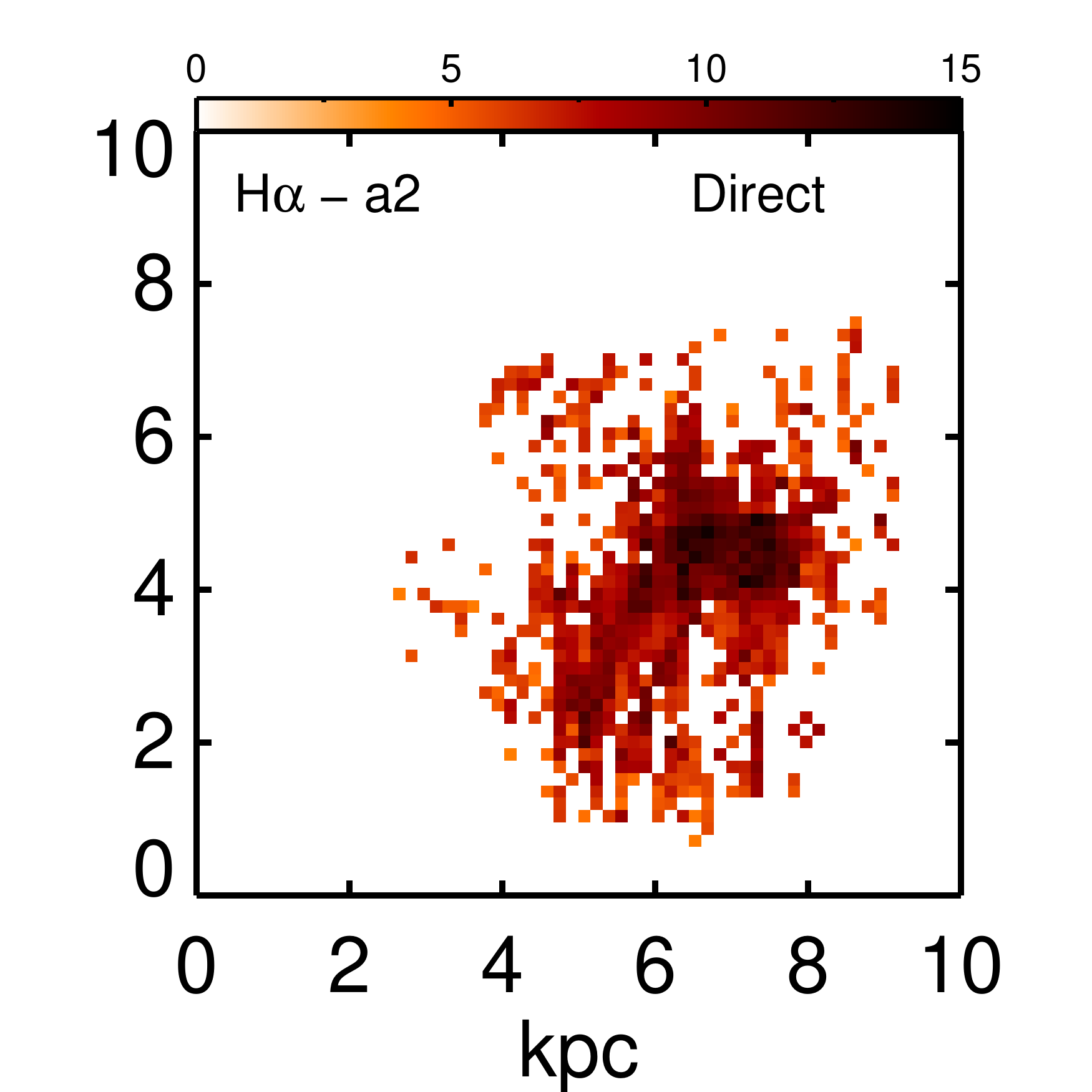}
\caption{}
\label{fig:fig2d}
\end{subfigure}
\caption{(a) Reconstruction of three wavelength channels with strong H$\alpha$ emission using the pixelated (upper panels) and direct (lower panels) methods of source reconstruction from the a1 datacube. Different values of {$\Delta$v} at the top of the panels represent the velocity offset from the systemic redshift of the source. The yellow curve shown in the middle panel represents caustics (regions of high magnification in the source plane) using the best fit lens model. The blue ellipses show the FWHM of the effective source plane PSF's at two different locations in the source plane. (b) Derived 2D source plane H$\alpha$ intensity maps in units of $10^{-16}$ erg s$^{-1}$ cm$^{-2}$ arcsec$^{-2}$ using the pixelated and direct method of source reconstruction for the a1 lensed image. The cyan ellipses refer to the clumps detected by the {\astrodendro} in the pixelated reconstruction of the a1 lensed image as described in Section~\ref{sec:discuss}. (c) and (d)  same as (a) and (b) respectively, but for the lensed image a2.}
\label{fig:fig2}
\end{figure*}

\section{Pixelated grid algorithm}\label{sec:method}
This work builds upon the observations of cswa128 and detailed lens modelling results from {\lenstool} presented in \cite{Sharma18} (hereafter Paper I). The target galaxy, lensed by a galaxy group at $z = 0.214$, is composed of two merging high magnification arcs a1 and a2 (Figure~\ref{fig:fig1}). The preliminary source reconstructions of the Integral Field Unit (IFU) observations with the OSIRIS
instrument on Keck I telescope are shown in Paper I using a traditional ray-tracing approach. In this letter, we revise the reconstructions using the pixelated technique to resolve clumps in the source-plane morphology of cswa128.

The source reconstruction in {\lenstool} was so far limited to simple parametric source profiles such as Gaussian or sersic light distributions. In reality, giant lensed arcs at high redshifts can have more complex surface brightness distributions (SBD). With the pixelated approach, we obtain a discretized SBD on a uniformly pixelized  source plane grid using a fixed mass profile of the lens. We use a Bayesian Monte Carlo Markov Chain (MCMC) algorithm called Massive Inference \citep[MassInf; part of the {\bayesys} package in {\lenstool}][]{Skilling98} to optimize the brightness of every pixel in the source-plane grid. The constraints come from the extended image plane distribution of the target galaxy.  A weighting map is also supplied to {\lenstool} that accounts for the uncertainty in the observed data. We ensure regularization or smoothness in the source profile by allowing every grid-based pixel to be defined by a Gaussian radial basis function (RBF) of a known size. The Bayesian MCMC algorithm samples the posterior distribution and outputs a number of reconstructed source frames and their corresponding image reconstructions. We create an average of these sampled frames and use it as the \textquotesingle best\textquotesingle \ reconstruction. A detailed description of the pixelated grid algorithm along with simulations used for testing the method will be provided in our follow-up paper Sharma et al (in prep). In summary, using the mock data on a variety of source profiles, we showed that the pixelated source modelling algorithm consistently results in significantly lower RMS errors (upto a factor of 5) than the traditional method of reconstruction. This letter focuses on demonstrating the strength of this algorithm in scientific applications through a case study.

\section{Application to cswa128}\label{sec:results}
We extract 14 wavelength channels in the vicinity of the H$\alpha$ emission line from the IFU data corresponding to a1 and a2 images in the lensed system. Then we reconstruct these channels in individual {\lenstool} runs using two different methods: direct reconstruction (or traditional ray-tracing) and pixelated grid algorithm. In terms of code, the source reconstruction is performed by the \textit{cleanlens} task in {\lenstool}. The direct method of reconstruction converts the image plane fluxes to the source plane with ray-tracing provided by the lens model that transfers the PSF to the source plane. We introduce the pixelated approach as a new feature of the \textit{cleanlens} task in the {\lenstool} software. The source plane resolution enabled by the lensing amplification is obtained after a sub-sampling of 5 on the source grid, resulting in a pixel scale of $0.\!\!^{\prime\prime}02$. 

The main inputs to the pixelated algorithm in {\lenstool}, in addition to the lens model best fit parameters, are the flux uncertainty, full width at half maximum (FWHM) of the Gaussian PSF in the image plane and FWHM of the Gaussian RBF in the source plane. To estimate the flux uncertainty in the data, we not only consider the variance of emission-free regions of the datacube, but also its correlation between pixels. We use the value of AO corrected PSF of $0.\!\!^{\prime\prime}12$ and $0.\!\!^{\prime\prime}15$ for IFU observations of a1 and a2 lensed images respectively. The size of the Gaussian RBF is defined as a factor of the sampling in the source plane to remove any dependencies of the source pixel size on the final reconstruction. Specifically, we test values of 1.5, 1.75, 2.0, 2.25, 2.5 as the ratios of the FWHM of the RBF to the sampling in the source plane. We note relatively higher image plane $\chi^{2}$ for RBF $> 2.0$ but do not find significant differences among the others. Therefore, we present the results using a RBF ratio of 1.5. 

For every pixel in the reconstructed source plane, we fit a Gaussian profile to the H$\alpha$ emission line for the a1 and a2 lensed images using a weighted $\chi2$ minimization procedure with the redshifts, the Gaussian FWHM and the flux of the emission line as free parameters (see Section 2.2.2 in Paper I for more details). The fitted spectra are weighted by a variance spectrum calculated using the reconstructed emission-free channels close to the redshifted H$\alpha$ wavelength. Figures~\ref{fig:fig2a} and~\ref{fig:fig2c} show snapshots of the reconstructed source frames around the redshifted H$\alpha$ emission line using the pixelated and direct approach for the a1 and a2 lensed images respectively. The corresponding 2D H$\alpha$ maps generated for both the lensed images are shown in Figure~\ref{fig:fig2b} and~\ref{fig:fig2d}. This figure illustrates the advantage of the pixelated approach in recovering the intrinsic shape of the source profile and in maximizing the SNR on the H$\alpha$ emission, especially in the highly distorted regions (clump IDs: 3 and 4) near the caustics. A clear distinction between the two approaches is more apparent for a1 than a2 because of the higher SNR in the observational data of a1 lensed image.

Quantitatively, the maximum linear resolution achieved in the direct reconstruction as derived from the FWHM of the projected image plane PSF in the direction of the greatest magnification varies between 25-290 pc. With the pixelated technique, we achieve a uniform resolution of $\sim 250$ pc derived from the size of the Gaussian RBF used to reconstruct the source plane. Even though the maximum linear resolution is higher with the direct approach, the irregular source plane PSF introduces uncertainties in the measurement of clump sizes, as demonstrated in the next section. 

\begin{figure*}
\centering
\includegraphics[scale = 0.54]{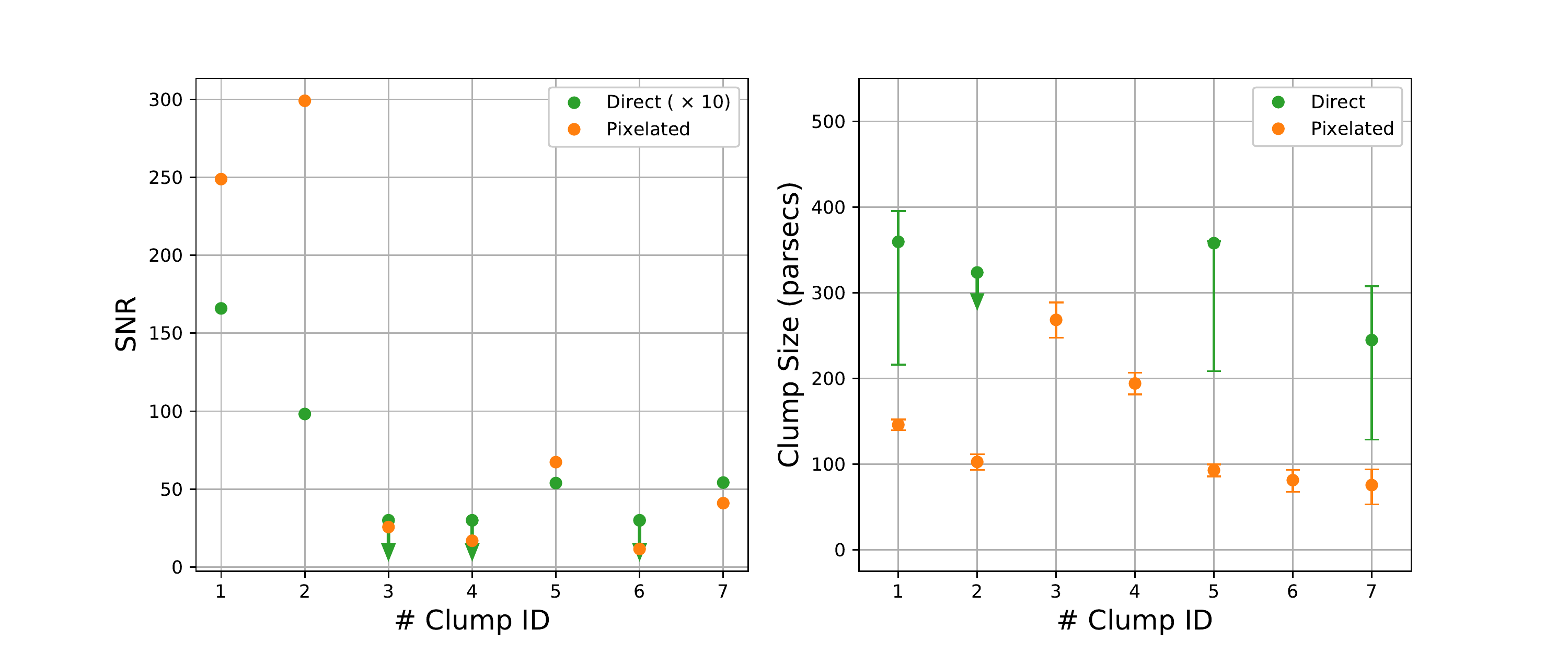}
\caption{Clump SNR and sizes: pixelated vs direct reconstruction. \textit{\bf Left:} SNR comparison for all the clumps detected in the source-plane reconstruction of H$\alpha$ maps as labelled in Figure~\ref{fig:fig2b}. SNR estimated for detections in the direct reconstruction (green data points) has been multiplied by a factor of 10 to exemplify the difference in values between the two methods. Only clump IDs 1, 2, 5 and 7 are detected in the direct reconstruction. For the other clumps, we plot the SNR as an upper limit, set at 3. \textit{\bf Right:} PSF corrected clump sizes characterized by Gaussian HWHM for detections in the H$\alpha$ maps obtained with the two methods of reconstruction. This figure highlights that the pixelated method not only increases the SNR of the clump detections by an order of magnitude, but also decreases the measured sizes and their uncertainty. }  
\label{fig:fig3}
\end{figure*}

\section{Clump Measurements}\label{sec:discuss}
In this section, we report the clump sizes measured in the derived source-plane H$\alpha$ profile of the a1 lensed image computed using both the methods of reconstruction. We do not consider a2 because of the low SNR in the observed datacube. We use the {\astrodendro}\footnote{http://www.dendrograms.org/} python package to compute hierarchical trees of structures called dendograms \citep{Ros08, Goodman09} and detect clumps in the H$\alpha$ maps. As compared to other non-hierarchical methods \citep[e.g. {\clumpfind},][]{Williams94}, {\astrodendro} prevents the formation of  \textquotesingle pathological\textquotesingle \ small clumps between two prominent clumps and allows structures with very different physical scales to be detected \citep{Goodman09}.

The algorithm selects clumps on the basis of three input parameters: (a) minimum value of flux that acts as a threshold to define clumps; we set it to 5$\sigma$ where $\sigma$ is the flux dispersion at a given source pixel (b) minimum difference in flux between two close structures for them to be considered separate; we set it to 10 \% following \cite{Oklo17} (c) minimum number of pixels in a clump, which is set to be 4 pixels corresponding to an effective radius of selected clumps to be greater than the source plane RBF size used for reconstruction. 

After computing the dendrograms, the {\astrodendro} package extracts the relevant parameters such as the location, surface area, effective major, minor axis and total fluxes of the identified structures. We find seven clumps in the pixelated reconstruction that satisfy the criteria we imposed. However, not all of these identified clumps are recovered using the traditional approach. We define the effective radius of a clump as the half width at half maximum (HWHM) derived from the geometric mean of the semi-major and semi-minor axis of the ellipse fitted to that clump in the source plane emission-line map. Additionally, we correct the effective radii of the clumps by subtracting the appropriate PSF HWHM calculated from either the seeing FWHM (direct reconstruction) or the RBF size (pixelated reconstruction) at the clump location in quadrature.

We use the fitted elliptical apertures in the pixelated reconstruction as a guide to measure the SNR of the individual clumps. The flux uncertainty and SNR of the clumps are based on the standard deviation of the total flux within a given aperture in the emission-line free regions of the source.

Figure~\ref{fig:fig3} compares the SNR and PSF corrected physical sizes (effective radii) of the clumps detected in the a1 source plane H$\alpha$ maps for both the reconstruction techniques. SNR of the clumps in the pixelated approach is higher than the traditional approach by almost a factor of 10. For non-detections, we report an upper limit of 3 on the SNR in the direct reconstruction. In particular, smaller clumps near the caustics (IDs: 3, 4 and 8 in Figure~\ref{fig:fig2}) are resolved with greater SNR using the pixelated reconstruction. 

As shown in Figure~\ref{fig:fig3}, the effective size estimates of the clumps are always lower using the pixelated method. For detections, where either of the effective axis estimate returned by the {\astrodendro} is smaller than the size of the source plane PSF at that location, we plot the upper limit at the value of the source plane PSF.

We use the jackknife method to calculate the uncertainties in the clump sizes derived using the pixelated approach. For each wavelength channel, we produce n realisations of the reconstructions by averaging randomly drawn samples returned by the MCMC sampler. Then, the H$\alpha$ emission line fitting is repeated n times and {\astrodendro} measurements are computed on each of these fitted H$\alpha$ maps. When necessary, the effective radii of the ellipses measured by the {\astrodendro} package are thresholded at the PSF size along a particular axis. Finally, the uncertainty is calculated as the standard deviation of the effective radii from all the random realizations in the jackknife tests.

\section{Conclusions}
In this work, we recover the physical sizes of H\,{\sc ii }\rm regions that probe star-formation in cswa128 using an automated forward modelling technique for source-plane reconstruction of lensed galaxies. The technique delenses a uniformly pixelized grid, deconvolved from the source plane PSF, using efficient Bayesian optimization. The case study of the pixelated technique on cswa128, as presented in this work, clearly shows that the approach is successful in minimizing the anisotropic distortion due to the PSF in the source plane, yielding more accurate source profiles. 

The power of the pixelated method is shown through (1) the larger number of clump detections (2) the higher SNR values in clump detections, some of which are undetected with the direct reconstruction, and (3) the recovery of the most distorted and faintest clumps near the caustics in the source plane. 
Studying how the star-forming clumps evolve with redshift is fundamental to galaxy evolution. The frontier of clump size measurements is being continuously pushed by gravitational lensing \citep[e.g.][]{Johnson17,Vanzella20}. Our pixelated technique provides a timely application to the upcoming IFS lensed galaxy studies using the James Webb Space telescope (JWST). This will be crucial in interpreting the star-formation processes at high redshifts and understanding galaxy assembly with unprecedented accuracy. 

The authors wish to recognize and acknowledge the very significant cultural role and reverence that the summit of Mauna Kea has always had within the indigenous Hawaiian community. We are grateful to the Keck Observatory staff for assistance with our observations. This research was supported by ASTRO 3D, through project number CE170100013. LJK acknowledges ARC Laureate Fellowship FL150100113. JR acknowledges support from the ERC Starting Grant 336736-CALENDS. SS thanks useful discussions with GEARS3D and CALENDS group. SS is also thankful to Surya Narayan Sahoo for technical help in producing figures of the paper. \\

\hspace{-0.6cm}\textbf{Data Availability:} The data underlying this article will be shared on reasonable request to the corresponding author.
\bibliographystyle{mn2e_v2}
\bibliography{bib_soniya}

\end{document}